\documentstyle[aps,prb,eqsecnum]{revtex}
\begin{document}
 
\title{Charge and spin dynamics in the one-dimensional $t$-$J_z$ and $t$-$J$
       models}
 
\author{Shu Zhang,\footnote{Present address: C.W. Costello \& Associates,
        Providence RI 02903} Michael Karbach, and Gerhard M\"uller}
 
\address{Department of Physics, The University of Rhode Island, Kingston,
         Rhode Island 02881-0817.}
 
\author{Joachim Stolze}
 
\address{Institut f\"ur Physik, Universit\"at Dortmund, 44221 Dortmund, Germany}
 
\date{\today}
\maketitle
%
%
\begin{abstract}
%
%
The impact of the spin-flip terms on the (static and dynamic) charge and spin
correlations in the Luttinger-liquid ground state of the one-dimensional (1D)
$t$-$J$ model is assessed by comparison with the same quantities in the 1D
$t$-$J_z$ model, where spin-flip terms are absent. We employ the recursion
method combined with a weak-coupling or a strong-coupling continued-fraction
analysis. At $J_z/t=0^+$ we use the Pfaffian representation of dynamic spin
correlations. The changing nature of the dynamically relevant charge and spin
excitations on approach of the transition to phase separation is investigated in
detail. At the transition point, the $t$-$J_z$ ground state has zero (static)
charge correlations and very short-ranged (static) spin correlations, whereas the
$t$-$J$ ground state is critical.  The $t$-$J_z$ charge excitations (but not the
spin excitations) at the transition have a single-mode nature, whereas charge
and spin excitations have a complicated structure in the $t$-$J$ model. A major
transformation of the $t$-$J$ spin excitations takes place between two distinct
regimes within the Luttinger-liquid phase, while the $t$-$J_z$ spin excitations
are found to change much more gradually. In the $t$-$J_z$ model, phase
separation is accompanied by N\'eel long-range order, caused by the condensation
of electron clusters with an already existing alternating up-down spin
configuration (topological long-range order). In the $t$-$J$ model, by contrast,
the spin-flip processes in the exchange coupling are responsible for continued
strong spin fluctuations (dominated by 2-spinon excitations) in the
phase-separated state.
\end{abstract}
 
\pacs{7.450Gb...}
\twocolumn
%
%
\section{Introduction}
%
%
At the heart of many phenomena in condensed matter physics is the interplay
between the charge and spin degrees of freedom of interacting electrons. The
impact of the magnetic ordering and fluctuations on the charge correlations or
the effect of the phase separation on the spin correlations, for example, are
important issues in the study of strongly correlated electron systems. One of
the simplest scenarios in which these questions can be formulated transparently
and investigated systematically comprises two successive approximations of the
Hubbard model with very strong on-site repulsion. They are known under the
names $t$-$J$ and $t$-$J_z$ models.\cite{ZR88}
 
Here we consider a one-dimensional (1D)
lattice.\cite{HM91,AW91,PS91,OLSA91,KY90,BBO91,THM95,YO91} In both models
the assumption is that the Hubbard on-site repulsion is so strong that double
occupancy of electrons on any site of the lattice may as well be prohibited
completely. This constraint is formally incorporated into the two models by
dressing the fermion operators of the standard hopping term with projection
operators:
\begin{equation}\label{I.1}
        H_t = -t\sum_{\sigma =\uparrow ,\downarrow }\sum_{l} \left\{
        \tilde c_{l,\sigma }^{\dagger }\tilde c_{l+1,\sigma }\ +
        \tilde c_{l+1,\sigma }^{\dagger }\tilde c_{l,\sigma }\right\}
\end{equation}
with $\tilde c_{l,\sigma } =c_{l,\sigma }(1-n_{l,-\sigma })$, $n_l=n_{l,\uparrow
}+n_{l,\downarrow }$, $n_{l,\sigma }=c_{l,\sigma }^{\dagger }c_{l,\sigma }$.
In the $t$-$J$ model the Hubbard interaction is further taken into account by an
isotropic antiferromagnetic exchange coupling between electrons on
nearest-neighbor sites:
\begin{equation}\label{I.2}
        H_{t\text{-}J} = H_t +
                J\sum_{l}\left\{{\bf S}_l\cdot {\bf S}_{l+1}-\frac 14 n_ln_{l+1}\right\}
\end{equation}
with $S_l^z=\frac 12(n_{l,\uparrow}-n_{l,\downarrow })$, $S_l^{+}=
\tilde c_{l,\uparrow }^{\dagger }\tilde c_{l,\downarrow }$, and $S_l^{-}=
\tilde c_{l,\downarrow}^{\dagger }\tilde c_{l,\uparrow }$.
In the $t$-$J_z$ model the isotropic exchange interaction is replaced by an
Ising coupling:
\begin{equation}\label{I.3}
        H_{t\text{-}J_z}=H_t +
        J_z \sum_{l} \left\{ S_l^z S_{l+1}^z - \frac 14 n_l n_{l+1} \right\}.
\end{equation}
 
The absence of spin-flip terms in $H_{t\text{-}J_z}$ introduces additional
invariants (not present in $H_{t\text{-}J}$) for the spin configurations of
eigenstates and thus alters the relationship between charge and spin
correlations considerably.  All results presented here will be for
one-quarter-filled bands ($N_e=N/2$ electrons on a lattice of $N$ sites).
 
For weak exchange coupling, both models have a Luttinger liquid ground
state. For stronger coupling, electron-hole phase separation sets in. Phase
separation is primarily a transition of the charge degrees of freedom. Here it
is driven by an interaction of the spin degrees of freedom, and it is
accompanied by a magnetic transition. The degree of spin ordering in the
phase-separated state depends on the presence ($t\text{-}J$) or absence
$(t\text{-}J_z)$ of spin-flip terms in the interaction.
 
Detailed information on the charge and spin fluctuations in $H_{t\text{-}J}$ and
$H_{t\text{-}J_z}$ is contained in the dynamic charge structure factor
$S_{nn}(q,\omega)$ and in the dynamic spin structure factor $S_{zz}(q,\omega)$,
i.e. in the quantity
\begin{equation}\label{I.4}
        S_{AA}(q,\omega )\equiv\int\limits_{-\infty}^{+\infty }dt
        e^{i\omega t}\langle A_q(t)A_{-q}\rangle,
\end{equation}
where $A_q$ stands for the  fluctuation operators
\begin{equation}\label{I.5}
        n_q   = N^{-\frac{1}{2}} \sum_l e^{-iql}n_l,\quad
        S_q^z = N^{-\frac{1}{2}} \sum_l e^{-iql}S_l^z.
\end{equation}
 
The degree of spin and charge ordering in the ground state is also reflected in
the equal-time charge correlation function $\langle n_ln_{l+m}\rangle$ and spin
correlation function $\langle S_l^zS_{l+m}^z \rangle$ and in their Fourier
transforms, the structure factors $S_{nn}(q)\equiv\langle n_q n_{-q} \rangle$
and $S_{zz}(q)\equiv \langle S_q^z S_{-q}^z\rangle$.
 
In the following we investigate the $T=0$ charge and spin fluctuations of the
two models $H_{t\text{-}J}$ and $H_{t\text{-}J_z}$ in three different regimes
with the calculational tools adapted to the situation: the limit of zero
exchange coupling (Sec. \ref{II}), the Luttinger liquid state
(Sec. \ref{Sec:LLS}), and the phase-separated state (Sec. \ref{Sec:PS}).
%
%
\section{Free lattice fermions}\label{II}
%
%
%
\subsection{Charge correlations and dynamics}
%
The tight-binding Hamiltonian (\ref{I.1}) has a highly spin-degenerate ground
state. The charge correlations are independent of the spin configurations and,
therefore, equivalent to those of a system of spinless lattice fermions,
\begin{equation}\label{II.1}
        H_t^\prime = -t \sum_l \left\{
                c_l^\dagger c_{l+1} + c_{l+1}^\dagger c_l
                                                   \right\}.
\end{equation}
This Hamiltonian has been well studied in the context of the 1D $s=1/2$ $XX$
model,
\begin{equation}\label{II.2}
        H_{XX} = -J_\perp \sum_l \left\{ S_l^x S_{l+1}^x + S_l^y S_{l+1}^y \right\},
\end{equation}
which, for $J_\perp = 2t$, becomes (\ref{II.1}) via Jordan-Wigner
transformation.\cite{LSM61,K62} The equal-time charge correlation function of
$H_t$ (or $H_t^\prime$) exhibits power-law decay,
\begin{equation}\label{II.3}
        \langle n_ln_{l+m}\rangle-\langle n_l\rangle\langle n_{l+m}\rangle =
        {{\cos (\pi m)-1} \over {2\pi ^2 m^2}},
\end{equation}
and the charge structure factor has the form
\begin{equation}\label{II.4}
        S_{nn}(q) - {N \over 4}\delta_{q,0} = {{|q|} \over {2\pi}}.
\end{equation}
The dynamic charge structure factor, which is equivalent to the $zz$ dynamic
spin structure factor of (\ref{II.2}) reads (for
$N\rightarrow\infty$):\cite{N67}
\begin{eqnarray}\label{II.5}
        S_{nn}(q,\omega )
&=&
        \pi^2\delta(q)\delta(\omega)
\nonumber\\ &+&
        \frac{2\Theta\biglb(\omega -2t\sin q\bigrb)
                   \Theta\biglb(4t\sin(q/2)-\omega\bigrb)}
                        {\sqrt{16t^2\sin ^2(q/2)-\omega^2}}.
\end{eqnarray}
%
\subsection{Spin correlations}
%
The charge-spin decoupling as is manifest in the product nature of the
ground-state wave functions of $H_{t\text{-}J_z}$ at $J_z/t=0^+$ and
$H_{t\text{-}J}$ at $J/t=0^+$ was shown to lead to a factorization in the spin
correlation function.\cite{PS91,PS90,OS90} We can write
\begin{equation}\label{II.6}
        \langle S_l^z S_{l+m}^z\rangle = \sum_{j=2}^{m+1} C(j-1) P(m,j),
\end{equation}
where $C(m)\equiv\langle S_l^zS_{l+m}^z\rangle_{LS}$ is the correlation function
in the ground state of a system of $N_e$ localized spins with antiferromagnetic
Heisenberg $(t$-$J)$ or Ising ($t$-$J_z$) coupling, and
\[
        P(m,j) \equiv \langle n_l n_{l+m}\delta_{j,N_m}\rangle, \quad
        N_m    \equiv \sum_{i=l}^{l+m} n_i
\]
is the probability of finding $j$ electrons on sites $l, l+1,\ldots,l+m$ with no
holes at the end points of the interval. This expression can be brought into the
form
\begin{eqnarray}\label{II.8}
        \langle S_l^z S_{l+m}^z\rangle = &&{-1 \over 4N_e} \sum_{k\neq 0}
                                                                         {S(k) \over \sin^2(k/2)}
\nonumber\\ && ~~
        \times[D_m(k)-2D_{m-1}(k)+D_{m-2}(k)],
\end{eqnarray}
\begin{eqnarray}
        S(k)   &=& \sum_{j=1}^{N_e}e^{ikj}C(j), \label{II.9}
\\
        D_m(k) &=& \left\langle \exp\left(-ik\sum_{l=0}^m n_l\right)\right\rangle,
        \nonumber
\end{eqnarray}
where $S(k)$ for $k=(2\pi/N_e)n$, $n=0,\ldots,N_e-1$ is the static structure
factor for the localized spins, and the $D_m(k)$ are many-fermion expectation
values, which are expressible as determinants of dimension $m+1$:\cite{PS91}
\[
        D_m(k)=\left|\delta_{ij}-{(1+e^{-ik}) \over 2N_e}
        {\sin[\pi(i-j)/2] \over \sin[\pi(i-j)/2N_e]}\right|_{i,j=0,\ldots,m}.
\]
In $H_{t\text{-}J_z}$ we have $C(m)\!=\!{1\over 4}(-1)^m$, i.e.
$S(k)\!=\!(N_e/4)\delta_{k,\pi}$, reflecting the (invariant) alternating up-down
sequence of successive electron spins. Expression (\ref{II.8}) can
then be evaluated in closed form:
 
\begin{mathletters}\label{II.12}
\begin{equation}
        \langle S_{l}^z S_{l + 2n }^z \rangle = \frac{(-1)^n}{2\pi^2}
        \prod_{i = 1}^{n-1} P_i^2,
\end{equation}
\begin{equation}\label{statzzodd}
        \langle S_{l}^z S_{l+2n+1}^z \rangle =
                -\frac{1}{2}\left(
                        \langle S_{l}^z S_{l+2n}^z \rangle +
            \langle S_{l}^z S_{l+2n+2}^z \rangle \right)
\end{equation}
\end{mathletters}
with
\[
        P_i = \frac{2}{\pi}\prod_{j=1}^i
                \left(1 - \frac{1}{4j^2} \right)^{-1}.
\]
The leading terms of the long-distance asymptotic expansion are\cite{note6}
\begin{eqnarray}\label{II.15}
        \langle S_{l}^zS_{l+m}^z\rangle_{t\text{-}J_z}
&&
    \stackrel{m\to\infty}{\longrightarrow}
        \frac{A^2}{4\sqrt{2}}\frac{1}{\sqrt{|m|}}
\nonumber\\ && \hspace*{-6mm}
        \times \left[
                \left(1-\frac{1}{8}\frac{1}{m^2}\right)\cos \frac{m\pi}{2}
        -\frac{1}{2m} \sin \frac{m\pi}{2}
        \right]
\end{eqnarray}
with $A = 2^{1/12}\exp[3\zeta^\prime(-1)] = 0.64500\ldots$. The structure of
$D_m(\pi)$ is very similar to that of the $xx$ spin correlation function of
$H_{XX}$.\cite{LSM61,M68,MPS83} Its leading asymptotic term has the form
$\langle S_l^x S_{l+m}^x\rangle_{XX} \sim (A^2/2\sqrt{2})m^{-1/2}$.
 
In $H_{t\text{-}J}$ the spin-flip terms weaken the spin correlations at
$J/t=0^+$.  The function $S(k)$ in (\ref{II.8}) is determined via (\ref{II.9})
by the spin correlation function of the 1D $s=1/2$ Heisenberg antiferromagnet
($XXX$ model).  Its leading asymptotic term reads\cite{SFS89} $C(m) \sim
\Gamma(-1)^mm^{-1}(\ln m)^{1/2}$ with amplitude $\Gamma\simeq 0.125(15)$ as
estimated from finite-chain data.\cite{note9} The leading asymptotic term of the
$t$-$J$ spin correlation function inferred from (\ref{II.8}) has the
form\cite{PS90}
\begin{equation}\label{II.16}
        \langle S_l^z S_{l+m}^z\rangle_{t\text{-}J}
        \sim \Gamma A^2\sqrt{2}{\cos(\pi m/2) \frac{(\ln m)^{1/2}}{m^{3/2}}}.
\end{equation}
The $t$-$J$ and $t$-$J_z$ spin structure factors $S_{zz}(q)$ inferred from the
results presented here will be presented and discussed in Sec. III.E.
 
For an intuitive understanding of the $q=\pi$ charge density wave in the ground
state at $J_z/t=0^{+}$ and $J/t=0^{+}$, we note that the hopping term opposes
electron clustering. In the absence of the exchange term, which favors
clustering of electrons with opposite spin, the hopping effectively causes an
electron repulsion. This is reflected in the power-law decay (\ref{II.3}) of
the charge correlation function, specifically in the term which oscillates with a
period equal to twice the lattice constant ($q=4k_F=\pi$). In this state, an
electron is more likely to have a hole next to it than another electron.
 
How does this affect the spin correlations? Recall that the ground state of
$H_{t\text{-}J_z}$ at $J_z/t=0^{+}$ is characterized by an (invariant)
alternating spin sequence. In a perfect electron cluster this sequence would
amount to saturated N\'eel ordering ($q=\pi $), but here it is destroyed by a
distribution of holes. Spin long-range order exists only in a topological
sense. However, some amount of actual spin ordering survives by virtue of the
effective electron repulsion in the form of the algebraically decaying term
(\ref{II.15}) in the spin correlation function with a wavelength equal to four
times the lattice constant ($q=2k_F=\pi/2$).
 
A similar argument obtains for the $t$-$J$ model. Since its ground state at
$J/t=0^{+}$ contains all spin sequences with $S_T^z=0$, not just the alternating
ones, the resulting $q=\pi /2$ oscillations (\ref{II.16}) in the spin
correlation function decay more rapidly than in the $t$-$J_z$ case.\cite{note1}
%
\subsection{Spin dynamics}
%
Expression (\ref{II.6}) cannot be generalized straightforwardly for the
calculation of {\em dynamic} spin correlations, the principal reason being that
the number of electrons between any two lattice sites is not invariant under
time evolution.  However, in the $t$-$J_z$ case we can determine the function
$\langle S_l^z(t)S_{l+m}^z\rangle$ on a slight detour. We use open boundary
conditions and write
\begin{equation}\label{II.17}
        S_{l}^z = - {1 \over 2}\sigma_L \prod_{i=1}^l (-1)^{n_i} n_l, \nonumber
\end{equation}
where $\sigma_L=\pm 1$ denotes the spin direction of the leftmost particle in
the chain, which {\em is} an invariant under time evolution. The time-dependent
two-spin correlation function of the open-ended $t$-$J_z$ chain is then related
to the following many-fermion correlation function:
\begin{eqnarray*}
        \langle S_l^z &(t)& S_{l+m}^z \rangle =
        {1\over 4}\left\langle n_l(t)\prod_{i=1}^l (-1)^{n_i(t)}\right.
        \left.\prod_{j=1}^{l+m} (-1)^{n_j} n_{l+m}\right\rangle
\nonumber \\ &=&
        \left\langle c_l^{\dag}(t) c_l(t) A_1(t) B_1(t)\right.\!
        A_2(t) B_2(t) \cdots A_l(t) B_l(t)
\nonumber \\ && ~~
        \times A_1 B_1 A_2 B_2 \cdots A_{l+m}\!
          \left.B_{l+m}c_{l+m}^{\dag}c_{l+m} \right\rangle
\end{eqnarray*}
with $ A_l\equiv c_l^{\dag} +c_l,\; B_l\equiv c_l^{\dag} -c_l$.  In order to
extract the bulk behavior of $\langle S_l^z(t)S_{l+m}^z \rangle$ from this
expression, we must choose both sites $l$ and $l+m$ sufficiently far from the
boundaries.
 
The numerical evaluation of this function via Pfaffians
shows\cite{SVM92,SNM95,note5} that the leading long-time asymptotic term
describes uniform power-law decay, $\langle S_l^z(t) S_{l+2n}^z\rangle \sim
t^{-1/2}$, for even distances and (more rapid) oscillatory power law decay, $
\langle S_l^z(t) S_{l+2n+1}^z\rangle \sim e^{-2it}t^{-\alpha},\; \alpha\gtrsim
1$, for odd distances.  Moreover, we have found compelling numerical evidence
that the relation (\ref{statzzodd}) can be generalized to time-dependent
correlation functions in the bulk limit $l\to\infty$.
 
From our data in conjunction with the long-distance asymptotic result
(\ref{II.15}) we predict that the leading term for large distances and long
times has the form\cite{note:t-energy}
\begin{equation}\label{tasymptjz}
        \langle S_l^z(t)S_{l+m}^z\rangle_{t\text{-}J_z}
        \sim {1 \over 4}{A^2/\sqrt{2} \over (m^2-4t^2)^{1/4}}
        \cos\frac{\pi m}{2}.
\end{equation}
The corresponding asymptotic result in the $XX$ model is well
established:\cite{MPS83,VT78}
\begin{equation}\label{II.19}
        \langle S_l^x(t)S_{l+m}^x\rangle_{XX} \sim {1 \over 4}{A^2\sqrt{2} \over
        (m^2-J_\perp^2t^2)^{1/4}}. \nonumber
\end{equation}
The asymptotic behavior (\ref{tasymptjz}) of the dynamic spin correlation
function implies that the dynamic spin structure factor has a divergent
infrared singularity at $q=\pi/2$: $S_{zz}(\pi/2,\omega)_{t\text{-}J_z} \sim
\omega^{-1/2}$. Further evidence for this singularity and for a corresponding
singularity in $S_{zz}(q,\omega)_{t\text{-}J}$ will be presented in
Sec.~III.F.
%
%
\section{Luttinger liquid state}\label{Sec:LLS}
%
%
Turning on the exchange interaction in $H_{t\text{-}J}$ and $H_{t\text{-}J_z}$,
which is attractive for electrons with unlike spins and zero otherwise, alters
the charge and spin correlations in the ground state gradually over the range of
stability of the Luttinger liquid state.
 
In the $t$-$J_z$ model, where successive electrons on the lattice have opposite
spins, the exchange coupling counteracts the effectively repulsive force of the
hopping term and thus gradually weakens the enhanced $q=\pi$ charge and
$q=\pi/2$ spin correlations. We shall see that the repulsive and attractive
forces reach a perfect balance at $J_z/t=4^-$. Here the distribution of
electrons (or holes) is completely random. All charge pair correlations vanish
identically and all spin pair correlations too, except those between
nearest-neighbor sites. This state marks the boundary of the Luttinger liquid
phase. At $J_z/t>4$ the attractive nature of the resulting force between
electrons produces new but different charge and spin correlations in the form of
charge long-range order at $q=0^+$ (phase separation) and spin long-range order
at $q=\pi $ (antiferromagnetism).
 
In the $t$-$J$ model the disordering and reordering tendencies are similar, but
the exchange interaction with spin-flip processes included is no longer
uniformly attractive. At no point in parameter space do the attractive and
repulsive forces cancel each other and produce a random distribution of
electrons. A sort of balance between these forces exists at $J/t=2$, which is
reflected in the observation\cite{YO91} that the ground-state is particularly
well represented by a Gutzwiller wave function at this coupling strength. Charge
and spin correlations exhibit power-law decay at the endpoint, $J/t\simeq 3.2$,
of the Luttinger liquid phase.  Here the attractive forces start to prevail on
account of sufficiently strong antiferromagnetic short-range correlations and
lead to phase separation, but the spin correlations continue to decay to zero
asymptotically at large distances.
 
One characteristic signature of a Luttinger liquid is the occurrence of infrared
singularities with interaction-dependent exponents in dynamic structure factors.
In the following we present direct evidence for interaction-dependent infrared
singularities in the dynamic charge and spin structure factors of
$H_{t\text{-}J_z}$ and $H_{t\text{-}J}$.  We employ the recursion
method\cite{rm} in combination with techniques of continued-fraction analysis
recently developed in the context of magnetic
insulators.\cite{VM94,VZSM94,VZMS95,FKMW96}
 
The recursion algorithm in the present context is based on an orthogonal
expansion of the wave function $|\Psi_q^A(t)\rangle \equiv A_q(-t)|\phi\rangle$
with $A_q$ as defined in (\ref{I.5}). It produces (after some intermediate
steps) a sequence of continued-fraction coefficients
$\Delta^A_1(q),\Delta^A_2(q),\ldots$ for the relaxation function,
\[
        c_0^{AA}(q,z) = \frac{1}{\displaystyle z + \frac{\Delta^{A}_1(q)}
        {\displaystyle
        z +  \frac{\Delta^{A}_2(q)}{\displaystyle z + \ldots }  }  }\;, \nonumber
\]
which is the Laplace transform of the symmetrized correlation function
$\Re\langle A_q(t)A_{-q}\rangle/\langle A_qA_{-q}\rangle$.  The $T=0$ dynamic
structure factor (\ref{I.4}) is then obtained via
\[
        S_{AA}(q,\omega) =  4\langle A_qA_{-q}\rangle\Theta(\omega)\lim
         \limits_{\varepsilon \rightarrow 0}
                \Re [c_{0}^{AA} (q, \varepsilon - i\omega)] \;. \nonumber
\]
 
For some aspects of this study, we benefit from the close relationship of the
two itinerant electron models $H_{t\text{-}J_z}$ and $H_{t\text{-}J}$ with the
1D $s=1/2$ $XXZ$ model,
\[
        H_{XXZ}=H_{XX}-J_\parallel\sum_l S_l^zS_{l+1}^z, \nonumber
\]
a model for localized electron spins. The equivalence of $H_{t\text{-}J_z}$ and
$H_{XXZ}$ for $J_\parallel=J_z/2$ and $J_\perp=2t$ was pointed out and used
before.\cite{BBO91,PS91} Depending on the boundary conditions, it can be
formulated as a homomorphism between eigenstates belonging to specific invariant
subspaces of the two models.  The mapping assigns to any up spin and down spin
in $H_{XXZ}$ an electron and a hole, respectively, in $H_{t\text{-}J_z}$. The
spin sequence of the electrons in the subspace of interest here is fixed, namely
alternatingly up and down. The importance of this mapping derives from the fact
that the ground state properties of $H_{XXZ}$ have been analyzed in great
detail.\cite{YY66,DG66,LP75}
 
The $T=0$ dynamic charge structure factor $S_{nn}(q,\omega )$ of
$H_{t\text{-}J_z}$ is thus equivalent to the $T=0$ dynamic spin structure factor
$S_{zz}(q,\omega)$ of $H_{XXZ}$ throughout the Luttinger liquid phase, and we
shall take advantage of the results from previous studies of $XXZ$ spin
dynamics.\cite{SSG82,BM82} The spin dynamics of $H_{t\text{-}J_z}$ is not
related to any known dynamical properties of $H_{XXZ}$.
%
\subsection{Charge structure factor}
%
Certain dominant features of the dynamic charge structure factor
$S_{nn}(q,\omega)$ are related to known properties of the static charge
structure factor.  Figure \ref{F1} displays finite-$N$ data of $S_{nn}(q)$ for
various coupling strengths in the Luttinger liquid phase of (a)
$H_{t\text{-}J_z}$ and (b) $H_{t\text{-}J}$.
 
The alignment of the data points on a sloped straight line in the free-electron
limit represents the exact result (\ref{II.4}), which is common to both models.
The persistent linear behavior at small $q$ for nonzero coupling reflects an
asymptotic term of the form $\sim A_0m^{-2}$ in the charge correlation function
$\langle n_ln_{l+m}\rangle$, while the progressive weakening of the cusp
singularity at $q=\pi$ reflects an asymptotic term of the form $\sim A_1\cos(\pi
m)/m^{\eta_\rho}$ with a coupling-dependent charge correlation exponent
$\eta_\rho$. For $H_{t\text{-}J_z}$ this exponent is exactly
known:\cite{LP75}
\begin{equation}\label{III.4}
        \eta _\rho = 2/[1-(2/\pi)\arcsin (J_z/4t)]\;.
\end{equation}
 
No exact result exists for the $t$-$J$ case, but the prediction is that the
charge correlation exponent varies over the same range of values,\cite{OLSA91}
i.e. between $\eta_\rho=2$ at $J/t=0$ and $\eta_\rho=\infty$ at $J/t\simeq 3.2$.
For $J/t \gtrsim 1$, the data in Fig.~\ref{F1}(b) indicate the presence of a
third cusp singularity in $S_{nn}(q)$, namely at $q=\pi/2$, which reflects the
third asymptotic term, $\sim A_2\cos(\pi m/2)/m^{1+\eta_\rho/4}$, predicted for
the $t$-$J$ charge correlations.\cite{note3} No corresponding singularity is
indicated in the data of Fig.~\ref{F1}(a), nor is any corresponding asymptotic
term predicted in the $XXZ$ spin correlations.
 
At the endpoint of the Luttinger liquid phase ($J_z/t=4$), the $t$-$J_z$
ground-state wave function has the form
\begin{eqnarray}\label{III.5}
        |\phi_0\rangle
&&
        =\sum_{1\leq l_1<l_2<\ldots<l_{N/2}\leq N}
        {N\choose N/2}^{-1/2}|l_1,\ldots ,l_{N/2}\rangle
\nonumber\\ &&~~~~~~~~~~~~~~
        \times \frac 1{\sqrt{2}}\left\{
        |\uparrow\downarrow\uparrow\ldots\rangle -
        |\downarrow\uparrow\downarrow\ldots\rangle\right\}\;,
\end{eqnarray}
where $|l_1,\ldots ,l_{N/2}\rangle$ specifies the variable charge positions.
The electrons are distributed completely at random on the lattice, while the
sequence of spin orientations is frozen in a perfect up-down pattern. This state
is non-degenerate for finite $N$, and its energy per site is $N$-independent:
$E_0/N\!=\!-t$. For $N\!\rightarrow\!\infty$, the $t\text{-}J_z$ charge
correlations disappear completely, $\langle n_l n_{l+m}\rangle-\langle
n_l\rangle \langle n_{l+m}\rangle=\delta_{m,0}/4$ as is indicated by the
finite-$N$ data for $J_z/t=4$ in Fig.~\ref{F1}(a): $S_{nn}(q) -
(N/4)\delta_{q,0}=[N/4(N-1)](1-\delta_{q,0})$. The $t$-$J$ charge correlations,
by contrast, seem to persist at $J/t\simeq 3.2$.
%
\subsection{Charge dynamics (weak-coupling regime)}
%
Expression (\ref{II.5}) for the $T=0$ dynamic charge structure factor
$S_{nn}(q,\omega)$ of $H_t$ is modified differently under the influence of a
$J_z$-type or a $J$-type exchange interaction.  Within the Luttinger liquid
phase we distinguish two regimes for the charge dynamics: a {\it weak-coupling}
regime and a {\it strong-coupling} regime as identified in the context of the
recursion method.\cite{VZMS95}
 
In the framework of weak-coupling approaches, the dynamically dominant
excitation spectrum of $S_{nn}(q,\omega)$ is confined to a continuum as in
(\ref{II.5}) but with modified boundaries and a rearranged spectral-weight
distribution. Moreover, a discrete branch of excitations appears outside the
continuum. A weak-coupling continued-fraction (WCCF) analysis for
$S_{nn}(\pi,\omega)$ of $H_{t\text{-}J}$ and, in disguise, also of
$H_{t\text{-}J_z}$, namely in the form of $S_{zz}(\pi,\omega)$ for $H_{XXZ}$ was
reported in Ref. \onlinecite{VZMS95}.  Without repeating any part of that
analysis we recall here those results which are important in the present
context.
 
The renormalized bandwidth $\omega_0$ of the dynamic charge structure factor
$S_{nn}(\pi,\omega)$ versus the coupling constant as obtained from a WCCF
analysis is shown in the main plot of Fig.~\ref{F2} for both the $t$-$J_z$ model
($\Box$) and the $t$-$J$ model ($\circ$).  In the $XXZ$ context, $\omega_0$ is
the bandwidth of the 2-spinon continuum, which is exactly known.\cite{DG66}
Translated into $t$-$J_z$ terms, the expression reads
\begin{equation}\label{III.8}
        \omega_0/2t=(\pi/\mu)\sin\mu, \quad \cos\mu=-J_z/4t
\end{equation}
and is represented by the solid line.  Comparison with our data confirms the
reliability of the WCCF analysis.
 
Our bandwidth data for the $t$-$J$ model can be compared with numerical results
of Ogata {\it et al.}\cite{OLSA91} for the charge velocity $v_c$ as derived from the
numerical analysis of finite chains.  The underlying assumption is that the
relation $\omega_0=2v_c$, which is exact in $H_{t\text{-}J_z}$, also holds for
the $H_{t\text{-}J}$ model.  The $t\text{-}J$ charge-velocity results of
Ref. \onlinecite{OLSA91} over the entire range of the Luttinger liquid phase are
shown as full circles connected by a dashed line in the inset. The solid line
represents the exact $t\text{-}J_z$ charge velocity $v_c=\omega_0/2$ with
$\omega_0$ from (\ref{III.8}).
 
The dashed line in the main plot is the $t$-$J$ bandwidth prediction inferred
from the data of Ref. \onlinecite{OLSA91}. It is in near perfect agreement with
the WCCF data ($\circ$). The open squares in the inset show the WCCF data over a
wider range of coupling strengths. The renormalized bandwidth $\omega_0$ will
shrink to zero at the endpoint of the Luttinger liquid phase, and the spectral
weight will gradually be transferred from the shrinking continuum to states of a
different nature at higher energies.
%
\subsection{Infrared exponent}
%
In the Luttinger liquid phase, the dynamic charge structure factor has
an infrared singularity with an exponent related to the charge correlation
exponent:
\begin{equation}\label{III.9}
        S_{nn}(\pi,\omega)\sim \omega^{\beta_\rho},\;\;
        \beta_\rho = \eta_\rho - 2\;.
\end{equation}
The WCCF analysis yields specific predictions for $\beta_\rho$ in both models.
Our results plotted versus coupling constant are shown in the inset to
Fig.~\ref{F3} for $H_{t\text{-}J_z}$ ($\Box$) and $H_{t\text{-}J}$
($\circ$). The solid line represents the exact $t$-$J_z$ result inferred from
(\ref{III.4}).
 
We observe that the WCCF prediction for the infrared exponent ($\Box$) rises
somewhat more slowly from zero with increasing coupling than the exact
result. The solid line in the main plot depicts the inverse square of the exact
$t$-$J_z$ correlation exponent (\ref{III.4}) over the entire range of the
Luttinger liquid phase. The open squares represent the WCCF data for
$2+\beta_\rho=\eta_\rho$ extended to stronger coupling.  For $H_{t\text{-}J}$
the correlation exponent is not exactly known.  The solid circles interpolated
by the dashed line represent the prediction for $\eta_\rho $ of Ogata et
al.\cite{OLSA91} based on a finite-size analysis.  The dashed line in the inset
is inferred from the same data. It agrees reasonably well with the WCCF data for
$\beta_\rho$ ($\circ$).
 
The solid and long-dashed curves in the main plot suggest the intriguing
possibility that the exponents $\eta_\rho$ of the two models have the same
dependence on the scaled coupling constants $J_z/J_z^{(c)}$ with $J_z^{(c)}=4t$
and $J/J^{(c)}$ with $J^{(c)}\simeq 3.2t$.  The short-dashed line represents the
exact $t$-$J_z$ result (\ref{III.4}) thus transcribed for $H_{t\text{-}J}$. Its
deviation from the data of Ogata {\it et al.} are very small throughout the
Luttinger liquid phase.
 
In Ref. \onlinecite{VZMS95} we carried out a WCCF reconstruction of the function
$S_{nn}(\pi,\omega)$ for the $t$-$J$ model and the $t$-$J_z$ model (alias $XXZ$
model).\cite{note2} The observed spectral-weight distributions of both models
consisted of a gapless continuum with a cusp-like infrared singularity
($\beta_\rho >0$), a shrinking bandwidth $(\omega_0/2t<2)$, and a lone discrete
state outside the continuum near its upper boundary.
%
\subsection{Charge dynamics (strong-coupling regime)}
%
What happens to the dynamic charge structure factor $S_{nn}(q,\omega)$ as the
exchange interaction is increased beyond the weak-coupling regime of the
Luttinger liquid phase?  For the $t$-$J_z$ case the answer can be inferred from
known results for the spin dynamics of $H_{XXZ}$.\cite{SSG82,BM82} The continuum
of charge excitations with sine-like boundaries
\[
        \epsilon_L(q)={\pi t\sin\mu \over \mu}|\sin q|, \quad
        \epsilon_U(q)=2\epsilon_L(q/2),
\]
continues to shrink to lower and lower energies, and discrete branches of
excitations
\[
        \epsilon_n(q)={2\pi t\sin\mu \over \mu\sin y_n}\sin{q \over 2}
        \sqrt{\sin^2{q \over 2}+\sin^2y_n\cos^2{q \over 2}}
\]
with $y_n=(\pi n/2\mu)(\pi-\mu)$ emerge successively at $\mu=\pi/(1+1/n)$ from
the upper continuum boundary.\cite{JKM73,BM82} All these excitations carry some
spectral weight, at least for finite $N$, but most of the spectral weight in
$S_{nn}(q,\omega)$ is transferred from the shrinking continuum to the top
branch, the one already present in the WCCF reconstruction.\cite{VZMS95}
 
At the endpoint of the Luttinger liquid phase $J_z/t=4$, the continuum states
have been replaced by a series of branches $\epsilon_n(q)=(2t/n)(1-\cos q)$,
$n=1,2,\ldots$, all the spectral weight is carried by the top branch $(n=1)$,
and the dynamic charge structure factor reduces to the single-mode form
\[
        S_{nn}(q,\omega)=\pi^2\delta(q)\delta(\omega)
        +{\pi \over 2}\delta\left(\omega-J_z\sin^2{q \over 2}\right).
\]
In the framework of the recursion method applied to the exact finite-size ground
state (\ref{III.5}), this simple result follows from a spontaneously terminating
continued fraction with coefficients $\Delta_1(q)=J_z^2\sin^4(q/2),
\Delta_2(q)=0$.
 
The dynamically relevant charge excitation spectrum of $H_{t\text{-}J}$, which
has an even more complex structure, will be presented in a separate study. In
this case, exact results exist only at one point $(J/t=2)$ in the strong
coupling regime.\cite{BBO91}
%
\subsection{Spin structure factor}
%
The long-distance asymptotic behavior of the $t$-$J$ spin correlation function
in the Luttinger liquid phase was predicted to be governed by two leading
power-law terms of the form\cite{HM91,AW91,PS91,OLSA91}
\begin{equation}\label{III.13}
        \langle S_l^zS_{l+m}^z\rangle_{t\text{-}J}\sim
        B_1\frac 1{m^2} + B_2\frac{\cos(\pi m/2)}{m^{\eta_\rho/4+1}}\;,
\end{equation}
where $\eta_\rho$ is the charge correlation exponent discussed previously. The
open circles in Fig.~\ref{F4}(a) depict the spin structure factor
$S_{zz}(q)_{t\text{-}J}$ for $J/t=0^+$ of a system with $N=56$ sites as inferred
via numerical Fourier transform from the results for the spin correlation
function presented in Sec. II. The two asymptotic terms of (\ref{III.13}) are
reflected, respectively, in the linear behavior at small $q$ and in the pointed
maximum at $q=\pi/2$. The latter turns into a square-root cusp as
$N\rightarrow\infty$.  The extrapolated maximum is
$S_{zz}(\pi/2)_{t\text{-}J}=0.28(1)$ (indicated by a $+$ symbol). The
extrapolated slope at $q=0$ is $S_{zz}(q)_{t\text{-}J}/q=0.0847(20)$.  The
observed smooth minimum at $q=\pi$ suggests that $S_{zz}(q)_{t\text{-}J}$,
unlike $S_{nn}(q)_{t\text{-}J}$, has no singularity there. The extrapolated
value is $S_{zz}(\pi)_{t\text{-}J}=0.127019(2)$.
 
The predictions of (\ref{III.13}) that the linear behavior in
$S_{zz}(q)_{t\text{-}J}$ at small $q$ persists throughout the Luttinger liquid
phase and that the cusp singularity at $q=\pi/2$ weakens with increasing $J/t$
and disappears at the onset of phase separation are consistent with our result
for $J/t=3.2$, plotted in Fig.~\ref{F4}(b).  The open circles suggest a smooth
curve which rises linearly from zero at $q=0$. The smooth extremum at $q=\pi$
has turned from a minimum at $J/t=0^+$ into a maximum at $J/t=3.2$.
 
The solid line in Fig.~\ref{F4}(a) represents $S_{zz}(q)_{t\text{-}J_z}$ for the
free-fermion case $J_z/t=0^+$ as obtained from Fourier transforming
(\ref{II.12}). It differs from the corresponding $t$-$J$ result ($\circ$) mainly
in three aspects: (i) the rise from zero at small $q$ is quadratic instead of
linear, reflecting non-singular behavior at $q=0$, i.e. the absence of a
non-oscillatory power-law asymptotic term in $\langle
S_l^zS_{l+m}^z\rangle_{t\text{-}J_z}$; (ii) the singularity at $q=\pi/2$ is
divergent: $\sim |q-\pi/2|^{-1/2}$; (iii) the smooth local minimum at $q=\pi$
has a slightly higher value, $S_{zz}(\pi)_{t\text{-}J_z}\simeq 0.129$.
 
Over the range of the Luttinger liquid phase, the asymptotic term in $\langle
S_l^zS_{l+m}^z\rangle_{t\text{-}J_z}$ which governs the singularity in
$S_{zz}(q)_{t\text{-}J_z}$ at $q=\pi/2$ is of the form $\sim~B_2\cos(\pi
m/2)/m^{\eta_\rho/4}$.  As in the $t$-$J$ case, the singularity weakens
gradually and then disappears at the transition point, $J_z/t=4$.  The
finite-$N$ result of $S_{zz}(q)_{t\text{-}J_z}$ at $J_z/t=4$, ($\bullet$) in
Fig.~\ref{F4}(b), indeed suggests a curve with no
singularities. This is confirmed by the exact result,
\begin{equation}\label{III.14}
        S_{zz}(q)_{t\text{-}J_z}={1 \over 8}(1-\cos q)\;,
\end{equation}
inferred from the exact ground-state wave function (\ref{III.5}) for
$N\rightarrow\infty$. It reflects a spin correlation function which vanishes for
all distances beyond nearest neighbors.
%
\subsection{Spin dynamics}
%
Under mild assumptions, which have been tested for $H_{t\text{-}J_z}$ at
$J_z/t=0^{+}$, the following properties of the dynamic spin structure factors
$S_{zz}(q,\omega)$ of $H_{t\text{-}J}$ or $H_{t\text{-}J_z}$ can be inferred
from the singularity structure of $S_{zz}(q)$: (i) The excitation spectrum in
$S_{zz}(q,\omega)$ is gapless at $q=\pi/2$.  (ii) The spectral-weight
distribution at the critical wave number $q=\pi/2$ has a singularity of the
form:
\[
        S_{zz}\left({\pi \over 2},\omega\right)_{t\text{-}J_z}\sim
        \omega^{{\eta_\rho\over 4}-2}\;,\;\;
        S_{zz}\left({\pi\over 2},\omega\right)_{t\text{-}J}
        \sim \omega^{{\eta_\rho\over 4}-1}\;.
\]
In the weak-coupling limit $(\eta_\rho =2)$, this yields $\sim \omega^{-3/2}$
for $H_{t\text{-}J_z}$ and $\sim \omega^{-1/2}$ for $H_{t\text{-}J}$. In both
cases, the infrared exponent increases with increasing coupling. A landmark
change in $S_{zz}(\pi,\omega)$ occurs at the point where the infrared exponent
switches sign (from negative to positive). In the $t$-$J_z$ case this happens
for $\eta_\rho =8$ and in the $t$-$J$ case for $\eta_\rho =4$.  According to the
data displayed in Fig.~\ref{F3}, this corresponds to the coupling strengths
$J_z/t=3.6955\ldots$ and $J/t\simeq 2.3$, respectively.
 
The dynamic spin structure factor $S_{zz}(q,\omega)_{t\text{-}J_z}$ as obtained
via the recursion method combined with a strong-coupling continued-fraction
(SCCF) analysis\cite{VM94,VZSM94} is plotted in Fig.~\ref{F5} as a continuous
function of $\omega$ and a discrete function of $q=2\pi m/N$, $m=0,\ldots ,N/2$
with $N=12$ for coupling strengths $J_z/t=0^{+},2,3,4$.  This function has a
non-generic $(q \leftrightarrow \pi-q)$ symmetry, which obtains for the
dynamically relevant excitation spectrum and for the line shapes, but not for
the integrated intensity.  In the weak-coupling limit, $J_z/t=0^{+}$, the
spectral weight in $S_{zz}(q,\omega)$ is dominated by fairly well defined
excitations at all wave numbers.  The dynamically relevant dispersion is $|\cos
q|$-like.
 
With $J_z/t$ increasing toward the endpoint of the Luttinger liquid phase, the
following changes can be observed in $S_{zz}(q,\omega )$: The peaks at $q\neq
\pi /2$ gradually grow in width and move toward lower frequencies.  The $|\cos
q|$-like dispersion of the peak positions stays largely intact, but the
amplitude shrinks steadily. The central peak at the critical wave number
$q=\pi/2$ starts out with large intensity and slowly weakens with increasing
coupling. Between $J_z/t=3$ and $J_z/t=4$, it turns rather quickly into a broad
peak, signaling the expected change in sign of the infrared exponent.
 
The dynamically relevant dispersion of the dominant spin fluctuations as
determined by the peak positions in our SCCF data for $S_{zz}(q,\omega)$ is
shown in Fig.~\ref{F6} for several values of $J_z/t.$ The linear initial rise
from zero at $q=\pi/2$ is typical of a Luttinger liquid.  The amplitude of the
$|\cos q|$-like dispersion decreases with increasing $J_z/t$ and approaches zero
at the transition to phase separation.  At the same time, the line shapes of
$S_{zz}(q,\omega)_{t\text{-}J_z}$ tend to broaden considerably.  These trends
are not shared with the $t$-$J$ spin excitations as we shall see.
 
The SCCF analysis indicates that the Luttinger liquid phase of the $t$-$J$ model
can be divided into two regimes with distinct spin dynamical properties.  For
coupling strengths $0<J/t\lesssim 1$, the function
$S_{zz}(q,\omega)_{t\text{-}J}$, which is plotted in Fig.~\ref{F8}, exhibits
some similarities with the corresponding $t$-$J_z$ results.  The main
commonality is a well-defined spin mode at not too small wave numbers with a
$|\cos q|$-like dispersion.  This dispersion is displayed in the main plot of
Fig.~\ref{F9} for different $J/t$-values within this first regime of the
Luttinger liquid phase.
 
However, even in the common features, the differences cannot be overlooked: (i)
The $(q\leftrightarrow \pi-q)$ symmetry in the line shapes of
$S_{zz}(q,\omega)_{t\text{-}J_z}$ is absent in $S_{zz}(q,\omega)_{t\text{-}J}$.
(ii) The amplitude of the $|\cos q|$-like dispersion grows with increasing
$J/t$, contrary to the trend observed in Fig.~\ref{F6} for the corresponding
$t$-$J_z$ spin dispersion.  (iii) The gradual upward shift of the peak position
in $S_{zz}(\pi,\omega )_{t\text{-}J}$ is accompanied by a significant increase
in line width (see inset to Fig.~\ref{F10}).  Over the range $0\leq J/t\lesssim
1.25$, the trend of the $q=\pi$ spin mode is opposite to what one expects under
the influence of an antiferromagnetic exchange interaction of increasing
strength.  (iv) The intensity of the central peak in
$S_{zz}(\pi/2,\omega)_{t\text{-}J}$ is considerably weaker than in in
$S_{zz}(\pi/2,\omega)_{t\text{-}J_z}$.  The peak turns shallow and disappears
quickly with increasing coupling (see Fig.~\ref{F10}, main plot).  This
observation is in accord with the proposed dependences of the infrared exponents
on the coupling constants.  (v) The linear dispersion of the dynamically
relevant spin excitations have markedly different slopes above and below the
critical wave number $q=\pi/2$ (Fig.~\ref{F9}, main plot).  At long wavelengths
the spectral weight in $S_{zz}(q,\omega)_{t\text{-}J}$ is concentrated at much
lower frequencies than in $S_{zz}(q,\omega)_{t\text{-}J_z}$.\cite{note8}
 
As the coupling strength increases past the value $J/t\simeq 0.75$, the spin
modes which dominate $S_{zz}(q,\omega)_{t\text{-}J}$ in the first regime of the
Luttinger liquid phase broaden rapidly and lose their distinctiveness. There is
a crossover region between the first and second regime, which roughly comprises
the coupling range $1\lesssim J/t\lesssim 2$.  Over that range, the spin dynamic
structure factor tends to be governed by complicated structures with rapidly
moving peaks.
 
At the end of the crossover region, a new type of spin mode with an entirely
different kind of dispersion has gained prominence in
$S_{zz}(q,\omega)_{t\text{-}J}$, and it stays dominant throughout the remainder
of the Luttinger liquid phase. This is illustrated in Fig.~\ref{F11} for three
$J/t$-values in the second regime of the Luttinger liquid phase.  The dispersion
of these new spin modes gradually evolves with increasing coupling strength as
shown in the inset to Fig.~\ref{F9}.  Note that the frequency has been rescaled
by $J$ both here and in Fig.~\ref{F11}.  At $J/t\lesssim 2.0$ the dispersion has
a smooth maximum at $q=\pi $ and seems to approach zero linearly as
$q\rightarrow 0$.  As $J/t$ increases toward the transition point, the peak
positions in $S_{zz}(q,\omega)_{t\text{-}J}$ gradually shift to lower values of
$\omega/J$, most rapidly at $q$ near $\pi$.
%
%
\section{Phase separation}\label{Sec:PS}
%
%
The transition from the Luttinger liquid phase to a phase-separated state in
$H_{t\text{-}J_z}$ takes place at $J_z/t=4$. The equivalent $XXZ$ model
undergoes a discontinuous transition to a state with ferromagnetic long-range
order at the corresponding parameter value ($J_{\parallel}/J_{\perp}=1$). The
ground state at the transition is non-critical and degenerate even for finite
$N$. The $XXZ$ order parameter, $\overline{M}=N^{-1}\sum_lS_l^z,$ commutes with
$H_{XXZ}$.
 
Notwithstanding the exact mapping, the transition of $H_{t\text{-}J_z}$ at
$J_z/t=4$ is of a different kind. Only one of the $N+1$ vectors which make up
the degenerate $XXZ$ ground state at $J_{\parallel}/J_{\perp }=1$ is contained
in the invariant subspace that also includes the $t$-$J_z$ ground state. The
other vectors correspond to $t$-$J_z$ states with different numbers $N_e$ of
electrons. The $t$-$J_z$ ground state at $J_z/t=4$ for fixed $N_e=N/2$ is
non-degenerate and represented by the wave function $|\phi_0\rangle$ as given in
(\ref{III.5}).
 
The fully phase-separated state as represented by the wave function
\begin{eqnarray}\label{IV.6}
        |\phi _1\rangle \equiv
&&
        \frac 1{\sqrt{2N}}\sum_{l_1=1}^N|l_1,l_1+1,\ldots,l_1+N/2-1\rangle
\nonumber\\ && ~~~~~~~~
        \times\left\{ |\uparrow \downarrow \uparrow \ldots \rangle
        \pm|\downarrow \uparrow \downarrow \ldots \rangle \right\}
\end{eqnarray}
has an energy expectation value at $J_z/t=4,\langle E_1\rangle=-t(N-2)$, which
exceeds the finite-$N$ ground-state energy, $E_0=-tN$, pertaining to
$|\phi_0\rangle$. However, by comparing the $J_z$-dependence of the energy
expectation values (per site) of the two wave functions $|\phi_0\rangle$ and
$|\phi_1\rangle$,
\begin{eqnarray*}
        \tilde{e}_0\equiv \frac{1}{N}\langle\phi_0|H_{t\text{-}J_z}|\phi_0\rangle
&=&
        -t -\frac{1}{2}\!\left(\frac{J_z}{4}-t\right)\!\left(1-\frac{1}{N-1}\right),
\\
        \tilde{e}_1\equiv \frac{1}{N}\langle\phi_1|H_{t\text{-}J_z}|\phi_1\rangle
&=&
        -\frac{J_z}{4}\left(1-\frac{2}{N}\right),
\end{eqnarray*}
in the vicinity of the transition,  $J_z/t=4(1+\epsilon)$, we obtain
\[
        \tilde{e}_0-\tilde{e}_1
                \stackrel{N\to\infty}{\longrightarrow}\frac{\epsilon}{2t},
\]
which implies that the two levels cross at $J_z/t=4$ in the infinite system.
 
The transition to phase separation in $H_{t\text{-}J_z}$ is characterized by
the charge and spin order parameters,
\[
        Q_\rho   = \frac{1}{N} \sum_{l=1}^Ne^{i2\pi l/N}n_l\;, \quad
        Q_\sigma = \frac{1}{N} \sum_{l=1}^Ne^{i\pi l }S_l^z\;.
\]
Neither operator commutes with $H_{t\text{-}J_z}$. The phase-separated state of
$H_{t\text{-}J_z}$ is characterized, for $N\rightarrow \infty $, by a broken
translational symmetry, $\langle Q_\rho \rangle \neq 0$, and a broken spin-flip
symmetry, $\langle Q_\sigma \rangle \neq 0$.
 
In the $t$-$J$ model, the transition to the phase-separated state, which takes
place at $J/t\simeq 3.2,$ produces charge long-range order, $\langle Q_\rho
\rangle \neq 0$, but is not accompanied by the onset of spin long-range order,
$\langle Q_\sigma \rangle =0$. The similarities in the charge correlations and
the differences in the spin correlations of the two models are evident in the
finite-size static charge and spin structure factors.
%
\subsection{Charge structure factor}
%
The vanishing charge correlations in the finite-size $t$-$J_z$ ground state at
the onset of phase separation ($J_z/t=4)$ is reflected in the flat charge
structure factor $S_{nn}(q)$ as shown in Fig.~11(a). The corresponding $t$-$J$
result for $J/t\simeq 3.2$ as shown in Fig.~11(b) indicates that correlated
charge fluctuations do exist at the transition.
 
With the exchange coupling increasing beyond the transition point, the charge
structure factors of the two models become more and more alike and reflect the
characteristic signature of phase separation. Phase separation is associated
with an enhancement of $S_{nn}(q)$ in the long-wavelength limit. Because of
charge conservation, this enhancement is manifest, in a finite system, not at
$q=0$ but at $q=2\pi/N$. It is conspicuously present in the data for couplings
$J_z/t=4.5$ and $J/t=3.5$, not far beyond the transition point.
 
The charge correlation function for the fully phase separated state, as
represented by the wave function (\ref{IV.6}), is a triangular
function,\cite{note4} $\langle n_ln_{l+m}\rangle =1/2-|m|/N,\; |m|\leq N/2$.
This translates into a charge structure factor of the form
\begin{equation}\label{IV.8}
        S_{nn}(q)={\frac N4}\delta _{q,0}+\frac{{1+\cos (Nq/2)}}{{N(1-\cos q)}}
                        (1-\delta _{q,0}),
\end{equation}
as shown (for $N=12$) by the full diamonds in Fig.~\ref{F12}. This function
vanishes for all wave numbers $q=2\pi l/N$ with even $l$ and increases
monotonically with decreasing odd $l$. The data in Fig.~\ref{F12} suggest that
the phase separation is nearly complete before the exchange coupling has reached
twice the value at the transition. In the $t\text{-}J_z$ case, we already know
that complete phase separation is established (for $N\to\infty$) right at the
transition.
%
\subsection{Spin structure factor}
%
The extremely short-ranged spin correlations in the $t\text{-}J_z$ ground state
(\ref{III.5}) for $N\rightarrow \infty $ are reflected by the static spin
structure factor (\ref{III.14}). For finite $N$ the spin correlations at
distances $|n|\geq 2$ do not vanish identically. An exponential decay is
observed instead with a correlation length that disappears as $N\rightarrow
\infty $. Hence the difference between (\ref{III.14}) and the finite-$N$ data
depicted in Fig.~12(a) ($\bullet$). The $t$-$J$ spin structure factor near the
transition $(J/t\simeq 3.2)$ has a similar $q$-dependence except at small $q$,
where it tends to zero linearly instead of quadratically.
 
Whereas the charge structure factors of the two models become more and more
alike as the exchange coupling increases in the phase-separated state
(Fig.~\ref{F12}), divergent trends are observed in the respective spin structure
factors, on account of the fact that the $t$-$J_z$ model supports spin
long-range order, and the $t$-$J$ model does not.
 
The fully phase-separated state of the $t$-$J_z$ model is at the same time fully
N\'{e}el ordered. The spin correlation function in the state (\ref{IV.6}) reads
$\langle S_l^zS_{l+m}^z\rangle=\frac{1}{4}(-1)^m(1/2-|m|/N),\: |m|\leq N/2$
and the corresponding spin structure factor has the form
\begin{equation}\label{IV.11}
        S_{zz}(q)= \frac{N}{16}\delta_{q,\pi}+
                        \frac{1-\cos [N(\pi-q)/2]}{4N[1-\cos (\pi -q)]}(1-\delta_{q,\pi}).
\end{equation}
The function (\ref{IV.11}) vanishes (for even $N/2$) at all wave numbers $q=2\pi
l/N$ with even $l$, just as (\ref{IV.8}) did. The exception is the wave number
$q=\pi $, where $S_{zz}(q)$ assumes its largest value.
 
The $t$-$J$ spin structure factor evolves quite differently in the presence of
increasing phase separation as is illustrated in Fig.~12(b). The electron
clustering produces in this case the Heisenberg antiferromagnet, whose ground
state is known to stay critical with respect to spin fluctuations. The spin
structure factor of that model is known to be a monotonically increasing
function of $q$, which grows linearly from zero at small $q$ and (for
$N\rightarrow \infty $) diverges logarithmically at $q=\pi$.\cite{SFS89}
%
\subsection{Spin dynamics ($t$-$J$ model)}
%
The charge long-range order in the phase-separated state freezes out the charge
fluctuations in both models, and the accompanying spin long-range order in the
$t$-$J_z$ model freezes out the spin fluctuations too. What remains strong are
the spin fluctuations in the $t$-$J$ model.
 
At the transition to phase separation ($J/t\simeq 3.2$), the $q=\pi $ spin mode
in $S_{zz}(q,\omega )_{t\text{-}J}$ does not go soft. However, the gradual
electron clustering tendency in conjunction with the continued strengthening of
the antiferromagnetic exchange interaction brings about a softening in frequency
and an enhancement in intensity of the order-parameter fluctuations associated
with N\'{e}el order. Both effects can be observed in the reconstructed dynamic
spin structure factors at $J/t=3.25,4.0,5.0$ as shown in Figs.~11(c), 13(a), and
13(b).
 
A close-up view of the gradual transformation of the $q=\pi $ mode is shown in
Fig.~14(a). For sufficiently strong exchange coupling, the function $S_{zz}(\pi
,\omega )_{t\text{-}J}$ will be characterized by a strong i.e.  nonintegrable
infrared divergence, $\sim \sqrt{-\ln\omega}/\omega$,\cite{BCK96} which
characterizes the order-parameter fluctuations of the 1D $s=1/2$ $XXX$
antiferromagnet.
 
Figure 14(b) shows the gradual change in line shape and shift in peak position
of the function $S_{zz}(\pi /2,\omega )_{t\text{-}J}$ in the phase-separated
state. The peak, which starts out relatively broad at the transition, shrinks in
width, loses somewhat in intensity, and moves to a higher frequency. For
$J/t\gtrsim 5.0$ it settles at $\omega /J\simeq \pi /2$ in agreement with the
lower boundary, $\omega_L(q)=(\pi J/2)|\sin q|$, at $q=\pi/2$ of the 2-spinon
continuum. The width has shrunk to a value consistent with the width of the
2-spinon continuum at that wave number.
 
In the inset to Fig.~14 we show the evolution of the dynamically relevant
dispersion for $S_{zz}(q,\omega )_{t\text{-}J}$ in the phase-separated state, as
determined by the peak positions of our data obtained via SCCF
reconstruction. The dashed line represents the exact lower threshold of the
2-spinon continuum. The shift of the peak positions in our data is directed
toward that asymptotic position at all wave numbers for sufficiently large
$J/t$.
%
%
\acknowledgments
%
%
This work was supported by the U.\ S. National Science Foundation, Grant
DMR-93-12252, and the Max-Kade Foundation. Computations were carried out on
supercomputers at the National Center for Supercomputing Applications,
University of Illinois at Urbana-Champaign.
%
%

\pagebreak
%
%
%
%
\begin{figure}
\caption[1]{Static charge structure factor at $T=0$ of (a) the $t$-$J_z$ model
        and (b) $t$-$J$ model in the Luttinger liquid phase.  Results extracted from
        the ground-state wave function determined numerically for a system of $N=12$
        sites.}\label{F1}
\end{figure}
 
\begin{figure}
\caption[2]{Main plot: Renormalized bandwidth of the dynamically relevant charge
        excitations in the weak-coupling regime of the Luttinger liquid phase of the
        $t$-$J$ and $t$-$J_z$ models.  Inset: Charge velocity in the two models over
        the full range of the Luttinger liquid phase.  The open symbols represent
        weak-coupling continued-fraction data and the solid lines represent the
        exact expression (\ref{III.8}). The full circles are finite-chain data from
        Ref. \onlinecite{OLSA91}.}  \label{F2}
\end{figure}
 
\begin{figure}
\caption[3]{Inset: Infrared exponent $\beta_\rho$ as defined by (\ref{III.9}) in
        the weak-coupling regime of the Luttinger liquid phase of the $t$-$J$ and
        $t$-$J_z$ models.  Main Plot: Inverse square of the charge correlation
        exponent for both models over the full range of the Luttinger liquid phase.
        The open symbols represent weak-coupling continued-fraction data, the solid
        lines represent the exact expression (\ref{III.4}), and the short-dashed
        line the same expression with $J/3.2t$ substituted for $J_z/4t$.  The full
        circles are the finite-chain data from Ref. \onlinecite{OLSA91}.}\label{F3}
\end{figure}
 
\begin{figure}
\caption[4]{Static spin structure factor at $T=0$ of the $t$-$J_z$ and $t$-$J$
        models (a) in the free-electron limit and (b) at the transition to phase
        separation.  The data for $J_z=0^+$ are calculated via numerical Fourier
        transform of expression (\ref{II.12}). The data for $J=0^+$ are derived
        from expression (\ref{II.8}) as explained in the text.  The remaining
        results are extracted from the ground-state wave function determined
        numerically for systems of $N=12$ sites.}\label{F4}
\end{figure}
 
\begin{figure}
\caption[5]{Dynamic spin structure factor $S_{zz}(q,\omega)$ at $T=0$ in the
        Luttinger liquid phase of the $t$-$J_z$ model.  The results for $t=1$ and
        four different values of $J_z$ are obtained via strong-coupling
        continued-fraction reconstruction from the coefficients
        $\Delta_1,\ldots,\Delta_6$ and an unbounded gap terminator
        (Refs. \onlinecite{VM94,VZSM94}). The $\Delta_k$'s are extracted from the
        ground-state wave function for a system of $N=12$ sites.}\label{F5}
\end{figure}
 
\begin{figure}
\caption[6]{Dynamically relevant dispersions of the excitations dominating the
        dynamic spin structure factor $S_{zz}(q,\omega)$ at $T=0$ for $t=1$ and
        different values of $J_z$ within the Luttinger liquid phase of the
        $t$-$J_z$ model. The symbols, which are smoothly interpolated by solid
        lines, represent the peak positions of results such as shown in
        Fig.~\ref{F5}.}  \label{F6}
\end{figure}
 
\begin{figure}
\caption[8]{Dynamic spin structure factor $S_{zz}(q,\omega)$ at $T=0$ for $t=1$
        and two values of $J$ in the first regime of the Luttinger liquid phase of
        the $t$-$J$ model. The results are obtained by the same method as those of
        Fig.~5.}\label{F8}
\end{figure}
 
\begin{figure}
\caption[9]{Dynamically relevant dispersions of the excitations dominating the
        dynamic spin structure factor $S_{zz}(q,\omega)$ at $T=0$ for $t=1$ and
        different values of $J$ in the first regime (main plot) and the second
        regime (inset) of the Luttinger liquid phase of the $t$-$J$ model.  The
        symbols, which are smoothly interpolated by solid lines, represent the peak
        position of results such as shown in Figs. \ref{F8} and
        \ref{F11}.}\label{F9}
\end{figure}
 
\begin{figure}
\caption[10]{Line shape at $q=\pi$ (inset) and $q=\pi/2$ (main plot) of the
        dynamic spin structure factor $S_{zz}(q,\omega)$ at $T=0$ for $t=1$ and
        various values of $J$ in the first regime of the Luttinger liquid phase of
        the $t$-$J$ model.  The results are obtained by the same method as
        those in Fig.~\ref{F5}.}\label{F10}
\end{figure}
 
\begin{figure}
\caption[11]{Dynamic spin structure factor $S_{zz}(q,\omega)$ at $T=0$ in the
        second regime of the Luttinger liquid phase of the $t$-$J$ model.  The
        results for $t=1$ and three different values of $J$ are obtained by
        the same method as those in Fig.~\ref{F5}.}\label{F11}
\end{figure}
 
\begin{figure}%
\caption[12]{Static charge structure factor at $T=0$ of (a) the $t$-$J_z$ model
        and (b) the $t$-$J$ model in the phase-separated state.  Results extracted
        from the ground-state wave function determined numerically for systems of
        $N=12$ sites.}\label{F12}
\end{figure}
 
\begin{figure}
\caption[13]{Static spin structure factor at $T=0$ of (a) the  $t$-$J_z$ model
        and (b) the $t$-$J$ model in the phase-separated state.  Results
        extracted from the ground-state wave function determined numerically for
        systems of $N=12$ sites.}\label{F13}
\end{figure}
 
\begin{figure}
\caption[14]{Dynamic spin structure factor $S_{zz}(q,\omega)$ at $T=0$ in the
        phase-separated state of the $t$-$J$ model.  The results for $t=1$ and two
        values of $J$ are obtained by the same method as those in
        Fig.~\ref{F5}.}\label{F14}
\end{figure}
 
\begin{figure}
\caption[15]{Line shape of the dynamic spin structure factor (a)
        $S_{zz}(\pi,\omega)$ and (b) $S_{zz}(\pi/2,\omega)$ of the $t$-$J$ model in
        the phase-separated state. The results for $t=1$ and various values of $J$
        are obtained by the same method as those in Fig.~\ref{F5}. Inset:
        Dynamically relevant dispersions of the excitations dominating the dynamic
        spin structure factor $S_{zz}(q,\omega)$ at $T=0$ for $t=1$ and different
        values of $J$ in the phase-separated state of the $t$-$J$ model.  The
        symbols which are smoothly interpolated by solid lines, represent the peak
        position of results such as shown in Fig.~\ref{F14}.}\label{F15}
\end{figure}
 

\begin{references}
\bibitem{note6} The first part of (\ref{II.15}) was previously given in
                Ref.\onlinecite{PS90}.
\bibitem{note9} We have fitted the asymptotic term to the data for
                $N_e=4,6,\ldots,28$ and extrapolated the sequence of
                amplitudes.
\bibitem{note1} What this simple argument does not explain is why the spin
                                correlations are stronger ($\sim m^{-1/2}$ or $m^{-3/2}$) than
                                the charge correlations ($m^{-2}$).
\bibitem{note5} References on the properties of Pfaffians and their use for the
                calculation of many-fermion correlations can be found, for
                example, in the Appendix of Ref. \onlinecite{SVM92}.
\bibitem{note:t-energy} We have set the energy unit of (\ref{I.1}) equal to
                                unity ($t=1$) here, in order to avoid confusion with the time
                                variable.
\bibitem{note3} This cusp was also observed at $J/t\gtrsim 1$ in the Monte Carlo
                                data of Ref. \onlinecite{AW91} for considerably larger systems.
\end{references}
\end{document}